\input harvmac.tex
\vskip 1.5in
\Title{\vbox{\baselineskip12pt
\hbox to \hsize{\hfill}
\hbox to \hsize{\hfill }}}
{\vbox{
	\centerline{\hbox{String Theory and Emergent $AdS$ Geometry
		}}\vskip 5pt
        \centerline{\hbox{in Higher Spin Field Theories 
		}} } }
\centerline{Dimitri Polyakov$^{}$\footnote{$^\dagger$}
{polyakov@sogang.ac.kr ;
twistorstring@gmail.com
}}
\medskip
\centerline{\it Center for Quantum Space-Time (CQUeST)$^{}$}
\centerline{\it Sogang University}
\centerline{\it Seoul 121-742, Korea}
\centerline{\it and}
\centerline{\it Institute for Information Transmission Problems (IITP)}
\centerline{\it Bolshoi Karetny per. 19/1}
\centerline{\it 127994 Moscow, Russia}
\vskip .3in

\centerline {\bf Abstract}

We analyze the Weyl invariance constraints on higher spin vertex
operators in open superstring theory describing massless higher spin 
gauge field
excitations in $d$-dimensional space-time. We show that these  constraints lead
to low-energy equations of motion for  higher spin fields in $AdS$
space, with the leading order $\beta$-function for
the higher spin fields producing Fronsdal's operator
in $AdS_{d+1}$, despite that the
higher spin vertex operators are originally
defined in flat background.
The correspondence between the $\beta$-function in string theory
and $AdS_{d+1}$ Fronsdal operators in space-time is
found to be  exact for
$d=4$, while for other space-time dimensions it requires 
modifications of manifest expressions for the 
higher spin vertex operators.
 We argue that the correspondence considered in this paper is the leading
order of more general isomorphism between
Vasiliev's equations and equations of motion of
extended open string field theory (OSFT),
 generalized to include the higher spin operators.

\Date{July 2013}

\vfill\eject
\lref\fronsdalsit{C. Fronsdal, Phys.Rev. D20 (1979) 848-856}
\lref\klebpol{S. Gubser, I. Klebanov, A. M. Polyakov, Phys.Lett. B428 (1998) 105-114}
\lref\hvs{M. Vasiliev, arXiv:1203.5554}
\lref\selfp{S. Lee, D. Polyakov, Phys.Rev. D85 (2012) 106014}
\lref\selfpp{D. Polyakov, Phys.Rev. D84 (2011) 126004} 
\lref\seung{S. Lee, D. Polyakov, Phys. Rev. D85 (2012) 106014}
\lref\fvf{E.S. Fradkin, M.A. Vasiliev, Nucl. Phys. B 291, 141 (1987)}
\lref\fvs{E.S. Fradkin, M.A. Vasiliev, Phys. Lett. B 189 (1987) 89}
\lref\vcubic{M. A. Vasiliev, Nucl. Phys. B862 (2012) 341-408}
\lref\vmaf{M. A. Vasiliev, Sov. J. Nucl. Phys. 32 (1980) 439,
Yad. Fiz. 32 (1980) 855}
\lref\vmas{V. E. Lopatin and M. A. Vasiliev, Mod. Phys. Lett. A 3 (1988) 257}
\lref\sagnottinew{A. Sagnotti, arXiv:1112.4285, J.Phys A46 (2013) 214006}
\lref\sagnottis{D. Francia, J. Mourad, A. Sagnotti,  Nucl. Phys. B804 (2008), 383-420}
\lref\taronnao{M. Taronna, arXiv:1005.3061}
\lref\taronnas{A. Sagnotti, M. Taronna, arXiv:1006.5242 ,
Nucl.Phys.B842:299-361,2011}
\lref\bts{E. Buchbinder, A. Tseytlin, JHEP 1008 (2010) 057}
\lref\vmat{E.S. Fradkin and M.A. Vasiliev, Mod. Phys. Lett. A 3 (1988) 2983}
\lref\vmafth{M. A. Vasiliev, Nucl. Phys. B 616 (2001) 106 }
\lref\hsaone{M. A. Vasiliev, Fortsch. Phys. 36 (1988) 33}
\lref\hsatwo{E.S. Fradkin and M.A. Vasiliev, Mod. Phys. Lett. A 3 (1988) 2983}
\lref\hsasix{E. Sezgin and P. Sundell, Nucl. Phys. B 634 (2002) 120 }
\lref\fronsdal{C. Fronsdal, Phys. Rev. D18 (1978) 3624}
\lref\coleman{ S. Coleman, J. Mandula, Phys. Rev. 159 (1967) 1251}
\lref\haag{R. Haag, J. Lopuszanski, M. Sohnius, Nucl. Phys B88 (1975)
257}
\lref\skvnew{N. Boulanger, D. Ponomarev, E. Skvortsov, M. Taronna, arXiv:1305.5180}
\lref\zhs{E. Skvortsov, J. Phys. A42 (2009), 385401}
\lref\giombi{S. Giombi, X. Yin, JHEP 1009 (2010) 115}
\lref\fradkin{E. Fradkin, M. Vasiliev, Phys. Lett. B189 (1987) 89}
\lref\skvortsov{E. Skvortsov, M. Vasiliev, Nucl.Phys.B756:117-147 (2006)}
\lref\bbd{F. Berends, G. Burgers, H. Van Dam ,Nucl.Phys. B260 (1985) 295}
\lref\mvd{L. Brink, R.Metsaev, M. Vasiliev, Nucl. Phys. B 586 (2000) 183}
\lref\klebanov{ I. Klebanov, A. M. Polyakov,
Phys.Lett.B550 (2002) 213-219}
\lref\spinself{D. Polyakov, Phys.Rev.D82:066005,2010}
\lref\spinselff{D. Polyakov,Phys.Rev.D83:046005,2011}
\lref\svv{E.D. Skvortsov, M.A. Vasiliev,
Nucl. Phys.B 756 (2006)117}
\lref\sez{E. Sezgin, P. Sundell, JHEP 0507 (2005) 044}
\lref\selfsw{D. Polyakov, J. Phys. A46 (2013) 214012}
\lref\vasnew{M. A. Vasiliev, arXiv:12126071 , Class.Quant. Grav. 30 (2013) 104006}
\lref\szw{A. Sen, B. Zwiebach, JHEP 0003 (2000) 002}

\centerline{\bf  1. Introduction}

Both string theory and higher spin gauge theories have been immensely active and 
fascinating fields for many years. These two cutting-edge fields are in fact deeply
connected to each other. At this point,  our understanding  of this connection  
is very far from being complete, still leaving many profound and conceptual
questions unanswered.
In the meantime the interplay between higher spins  and strings
appears of crucial relevance to fundamental
questions such as underlying reasons behind AdS/CFT conjecture ~{\klebanov, \sez},
holography principle, origin of space-time geometry and others.
There exists a number of examples linking string and higher spin dynamics.
 It is well-known that massless higher spin modes appear in the tensionless limit
of string theory as the massive  vertex operators carrying  spin $s\sim{m^2}$ (where $m$ is the mass).
 This correspondence has been explored in a number of insightful papers
(e.g. see  for ~{\sagnottinew, \taronnas} for some reviews). This approach has many obvious advantages
(it is, in principle, straightforward to construct higher spin vertex operators
in the massive sector both in bosonic and superstring theory), however
it faces a number of  difficulties as well, many of them related to the fact that,
 in general, tensionless limit of string theory is the difficult one to describe.
In particular, it seems hard to recover the full set of Stuckelberg symmetries
when the vertex operators technically become massless, as $\alpha^\prime\rightarrow\infty$.
The vertex operators constructed in this approach can be used to describe
metric-like Fronsdal fields (rather than the frame-like gauge fields in 
Vasiliev's theory ~{\fvf, \fvs, \vmaf, \vmas, \vmat, \vmafth}),
therefore  fundamental space-time symmetries, related to higher-spin currents are not manifest
in this approach . Moreover, while this approach allows to understand the structure
of vertex operators in flat space-time, it is known to be difficult to extend it to the $AdS$
case (e.g. ~{\sagnottis}) since straightforward quantization of strings in $AdS$ background is not
known beyond the semiclssical limit ~{\bts}. At the same time,
$AdS$ geometry appears to be  pertinent and crucial ingredient in constructing consistenty
interacting higher spin theories. 
Apart from crucial relevance to AdS/CFT correspondence ~{\klebanov, \sez},
understanding  higher spin dynamics in $AdS$ space is of especial interest
to us since, it is the
$AdS$ geometry which circumvents the limitations of Coleman-Mandula's theorem
{\coleman, \haag},
leaving  possibility of constructing consistent higher spin interactions
at all orders, following the Vasiliev's equations ~{\giombi}.

In our previous papers we shown that, apart from the tensionless limit, higher spin vertex operators
describing emissions of higher spin gauge fields by an open string, also
can be constructed at an arbitrary tension value but at non-canonical
ghost pictures ~{\spinself, \spinselff}. 
These operators are related to global symmetries
present in RNS superstring theory , including hidden AdS isometries and their
higher spin extensions, that can be classified using the formalism of ghost 
cohomologies
 ~{\spinselff}. The generators inducing these symmetries do not mix with 
standard Poincare generators,
in this sense describing ``symmetries of different world'',  
co-existing with our world within
``larger'' superstring theory, that includes 
picture-dependent operators existing at 
nonzero ghost numbers,
which ghost dependence cannot be removed by picture-changing
(these states and related symmetries, however, have no effect on 
standard string perturbation theory)

 The hidden symmetry generators
can be conveniently classified in terms of ghost cohomologies $H_n$.
For the sake of completeness 
we briefly remind the definition of $H_n$ which properties were analyzed
in a number of previous works (e.g. see ~{\spinself, \spinselff})
For each positive $n>0$ $H_n$ is defined as a set of physical 
(BRST closed and BRST nontrivial)
vertex operators existing at minimal positive picture $n$ and above, annihilated
by inverse picture-changing transformation at minimal picture $n$
(with the picture transformations above the picture n generated by usual direct and inverse picture changings).
For each negative $n\leq{-3}$ and below $H_n$ is 
 is defined as a set of physical 
vertex operators existing at minimal positive picture $n$ and below (i.e. $n-1$, $n-2$,  etc) 
annihilated
by direct picture-changing transformation at minimal negative picture $n$
(with the picture transformations above the picture n generated by usual direct and inverse picture changings).
The cohomologies of positive and negative orders are isomorphic according to
$H_{n}\sim{H_{-n-2}}(n\geq{1})$. Also, $H_0$ by definition consists of all picture-independent operators
(existing at all picture representations) while $H_{-1}$ and $H_{-2}$ are empty.
Thus all conventional string theory operators (such as a photon, a graviton or Poincare generators)
are the elements of $H_0$ or $H_0\otimes{H_0}$.  The generators
inducing $AdS$ transvections in the larger string theory are the elements of $H_1\sim{H_{-3}}$
while the massless closed string vertex operator of spin 2 bilinear in transvections:
$H_1\otimes{H_1}\sim{H_{-3}}\otimes{H_{-3}}$ describes gravitational fluctuations around
the AdS vacuum (see below). Massless  open string operators of spin $s\geq{3}$
describing frame-like gauge fields in Vasiliev's theory
are the elements of $H_{s-2}\sim{H_{-s}}$; their explicit construction will be given below.
The fusion rules describing operator products between the vertices
of different $H_s$ have the  same structure as the higher spin algebras in $AdS$ space;
in other words the OPE algebra in the larger string theory constitutes one (and very convenient)
realisation of $AdS$  higher spin algebras.

Given the  global symmetry generators, it is then straightforward to construct the appropriate 
vertex operators in open and closed string theories  describing emissions of massless particles of various spins
(with the open string physical vertex operators being objects linear in the symmetry generators,
while the closed string operators are the objects bilinear in the symmetry generators.
The purpose of this work is to analyze how AdS geometry emerges in the $\beta$-function equations
 for the massless higher spin modes in RNS theory. In the leading order, the Weyl invariance
constraints on the higher spin vertex operators lead to low-energy equations of motion for
massless higher spin fields  defined by the Fronsdal operator in AdS space-time.
The AdS structure of the Fronsdal operator (with the appropriate mass-like terms) emerges despite the fact
that the higher spin vertex operators are initially defined in the flat background in RNS theory.
The appearance of the AdS geometry is directly related to the ghost cohomology structure of the higher spin vertices
and is detected through the off-shell analysis of the $2d$ scale invariance of the vertex operators for higher spins.
It is crucial that, in order to see the emergent $AdS$ geometry one must go off-shell, e.g. to analyze
the scale invariance of the operators in $2+\epsilon$ dimensions so that the
trace $T_{z{\bar{z}}}$ of the  stress-energy tensor generating $2d$ Weyl transformations is no longer identically zero.
Namely, it is the off-shell analysis of the operators at nonzero $H_{n}$
that allows to catch  cosmological type terms in low energy equations of motion  
while the on-shell constraints on the operators  (such as BRST conditions)  do not detect them,
only leading to standard Pauli-Fierz equations for massless higher spins in flat space.
This is a strong hint that the, from the string-theoretic
point of view, the appropriate framework to analyze the higher spin interactions is the off-shell
theory, i.e. string field theory, with the SFT equations of motion:
$Q\Psi=\Psi\star\Psi$ related to Vasiliev's equations in unfolding formalism. It is important
to stress, however, that Vasiliev's equations must be related to the enlarged, rather than ordinary SFT, with 
 the string field $\Psi$ extended to higher ghost cohomologies. Higher spin interactions
in AdS should then be deduced from the off-shell string field theory computations involving
higher spin vertex operators for Vasiliev's frame-like fields on the worldsheet boundary, with
the appropriate insertions of $T_{z{\bar{z}}}$ in the bulk  controlling the cosmological constant dependence.
The rest of this paper is organized as follows.
In the next section, we review the hidden AdS isometries in RNS string theory, construction
of vertex operator in $H_1\otimes{H_1}$ based on these isometries,
describing gravitational fluctuations around underlying AdS background
and appearance of the cosmological term in its beta-function
as a result of off-shell scale-invariance condition.
Next, we analyze the Weyl invariance of higher spin operators for
massless spin $s$ fields in Vasiliev's formalism, constructed in  $H_{s-2}\sim{H_{-s}}$
In this work we mostly limit ourselves to the peculiar case of the higher spins
that are polarized and propagating along the $AdS$ boundary (which nevertheless is the
limit relevant for holography) with also making some comments regarding the bulk-dependent
case. The Weyl invariance, in the leading order, leads to the low-energy equations of motion 
for the higher spins, determined by the Fronsdal's operator in AdS space.
In the concluding section we outline the higher order extension of this calculation
(currently in progress) in order to establish isomorphism between higher spin vertices
in AdS space and off-shell amplitudes 
in extended string field theory with $T_{z{\bar{z}}}$-insertions.The ultimate aim of this
program is to explore conjectured isomorphism between equations of extended SFT and
Vasiliev's equations that describe higher spin interactions in unfolded approach. 

\centerline{\bf 2. Hidden AdS Isometries and Gravitons in AdS}

The starting point is RNS  superstring theory in flat d-dimensional space-time,
with the action given by:
\eqn\grav{\eqalign{
S_{RNS}=S_{matter}+S_{bc}+S_{\beta\gamma}+S_{Liouville}
\cr
S_{matter}=-{1\over{4\pi}}\int{d^2z}(\partial{X_m}\bar\partial{X^m}
+\psi_m\bar\partial\psi^m+{\bar\psi}_m\partial{\bar\psi}^m)
\cr
S_{bc}={1\over{2\pi}}\int{d^2z}(b\bar\partial{c}+{\bar{b}}\partial
{\bar{c}})
\cr
S_{\beta\gamma}={1\over{2\pi}}\int{d^2z}(\beta\bar\partial\gamma
+\bar\beta\partial{\bar\gamma})
\cr
S_{Liouville}=-{1\over{4\pi}}\int{d^2z}(\partial\varphi\bar\partial\varphi
+\bar\partial\lambda\lambda
+\partial\bar\lambda\bar\lambda
+\mu_0{e^{B\varphi}}(\lambda\bar\lambda+F))
}}
where 
 $X^m(m=0,...{d-1})$ are the space-time coordinates and $\psi^m$ are their 
worldsheet superpartners, $b=e^{-\sigma},c=e^\sigma$ 
are reparametrization ghosts,
$\gamma=e^{\phi-\chi}\equiv{e^\phi\eta}$ abd $\beta=e^{\chi-\phi}\partial\chi\equiv\partial\xi{e^{-\phi}}$
aresuperconformal ghosts,
$\varphi,\lambda, F$ are components of super Liouville field
and the Liouville background charge is
$Q=B+B^{-1}={\sqrt{{{9-d}\over2}}}$.
The action (1) is obviously invariant under global Poincare symmetries
generated by
\eqn\grav{\eqalign{P^m=\oint{dz}\partial{X^m}(z)\cr
P^{mn}=\oint{dz}({\partial}X^{\lbrack{m}}\psi^{n\rbrack}+\psi^m\psi^n)}}
The standard physical vertex operators in RNS superstring theory are 
the objests that are
the elements of $H_0$, linear in Poincare generators $P$ 
(for open strings) or bilinear
(for closed strings), up to multiplication by the exponent field
 ${\sim}e^{ipX}$.
For example, the photon operator
is $V_m\sim\oint{dz}\Pi_m{e^{ipX}}$ and the graviton is 
$V_{mn}\sim\int{d^2z}\Pi_m{\bar\Pi}_n{e^{ipX}}(z,{\bar{z}})$.
where $\Pi_m=\partial{X_m}+i(p\psi)\psi_m$ at picture 0, 
$\Pi_m=e^{-\phi}\psi_m$ at picture $-1$ etc.
The crucial point is that, 
apart from obvious Poincare symmetries of flat space-time, 
the action (1) 
also has nonlinear global symmetries realising hidden $AdS$ 
isometry algebra.
Namely, as a warm-up example, it is straightforward to
 check the invariance of (1) under
\eqn\grav{\eqalign{
\delta_\alpha{X_m}=\alpha(\partial(e^\phi\phi_m)+2e^\phi\partial\phi_m)\cr
\delta_\alpha{\psi_m}=-\alpha(e^\phi\partial^2{X_m}
+2\partial(e^\phi\partial{X_m}))\cr
\delta_\alpha{\gamma}=\alpha{e^\phi}(\psi_m\partial^2{X^m}
-2\partial\psi_m\partial{X^m})\cr
\delta_\alpha{b}=\delta_\alpha{c}=\delta_\alpha\beta=0}}

as well as  under the dual version of these transformations,
given by replacing $\phi\rightarrow{-3\phi}$
 in the transformation laws for $X$ and $\psi$,
vanishing variations of $b,c$ and $\gamma$ ghosts and
the transformation of the $\beta$ ghost given by
\eqn\grav{\eqalign{\delta\beta=\partial\xi{e^{-4\phi}}
\sum_{k=0}^2P^{(k)}_{-3\phi}\partial^{(2-k)}F_{{5\over2}}}}
where $\alpha$ is global bosonic infinitezimal parameter, 
the polynomials $P_f^{(n)}=e^{-f(z)}{{d^n}\over{dz^n}}e^{f(z)}$ 
are the conformal weight $n$ operators
if $f(z)$ is linear in the ghost fields $\phi$,$\chi$ and $\sigma$ and 
$F_{{5\over2}}$ is dimension
${5\over2}$ primary field:
$F_{{5\over2}}=\psi_m\partial^2{X^m}-2\partial\psi_m\partial{X^m}$.
The generators of (3), (4) are easily constructed to be given by
\eqn\grav{T^{(+1)}=\oint{dz}e^\phi{F_{{5\over2}}}(z)}
for (3) and
\eqn\grav{T^{(-3)}=\oint{dz}e^{-3\phi}{F_{{5\over2}}}(z)}
for (4). The operator (6) is BRST invariant and nontrivial while the operator
(5) is not, as it doesn't commute with the supercurrent terms of $Q_{brst}$.
In order to make (5) BRST-invariant one has to modify it with 
$b-c$ ghost dependent
terms according to the homotopy $K$-transformation 
$T\rightarrow{L}={K}\circ{T}$
where, in general, $T$ is an operator given by an integral
of dimension 1 primary field $V$
, not commuting with $Q_{brst}$: 
$$T=\oint{dz}V(z)$$ 
and  the transformation is defined as
\eqn\grav{\eqalign{
K{\circ}T=T
+{{(-1)^N}\over{N!}}
\oint{{dz}\over{2i\pi}}(z-w)^N:K\partial^N{W}:(z)
\cr
+{1\over{{N!}}}\oint{{dz}\over{2i\pi}}\partial_z^{N+1}{\lbrack}
(z-w)^N{K}(z)\rbrack{K}\lbrace{Q_{brst}},U\rbrace
}}
where $w$ is some arbitrary point on the worldsheet,
$U$ and $W$ are the operators defined according 
to
\eqn\grav{\eqalign{
\lbrack{Q_{brst}},V(z)\rbrack=\partial{U}(z)+W(z)
}}

\eqn\grav{\eqalign{
K=ce^{2\chi-2\phi}
}}
is the homotopy operator satisfying
$${\lbrace}Q_{brst},K{\rbrace}=1$$
and $N$ is the leading order of the operator product
\eqn\grav{\eqalign{
K(z_1)W(z_2)\sim{(z_1-z_2)^N}Y(z_2)
+O((z_1-z_2)^{N+1})
}}

In case of the symmetry  generator (5) we have $N=2$.
It is straightforward to check that, with the definition (7),
the operator $K\circ{T}$ is BRST-invariant.
The homotopy transformation (7) is straightforward to generalize 
for closed string operators, multiplying it by
 antiholomorphic 
transformation, so that the invariant
closed string operator is $K{\bar{K}}\circ\int{d^2z}{...}$.
The important property of the $K$-transformation
is the homomorphism relation preserving the OPE structure constants so that,
up to BRST-exact terms, the OPE structure constants
of BRST-invariant operators $K\circ{T_1}$ and $K\circ{T_2}$ can be read off
the OPE of the non-invariant operators $T_1$ and $T_2$, with the appropriate
$K$-transform (7) of the right-hand side  (see ~{\spinselff} for the proof).
Given (6)-(10), the dual symmetry generators $T^{(-3)}$ and $K{\circ}T^{(+1)}$
belong to isomorphic cohomologies $H_{-3}$ and $H_1$
(note that both $T^{(+1)}$ and 
 $K{\circ}T^{(+1)}$ generate global symmetries in space-time;
however,  while the non-invariant operator  $T^{(+1)}$ generates
 the symmetry transformations (3) that do not involve the ghost fields
$b,c$ and $\beta$, the invariant operator
 $K{\circ}T^{(+1)}$ generates the extended (complete) version of
 (3) which involves all the ghost fields.
Given the definitions (5), (7), the extended space-time
transformations are straightforward to construct; we will not
present their manifest form here for the sake of brevity.
It shall be sufficient to note that, due to the homomorphism property
~{\spinselff}, the $K$-transformed symmetry generators satisfy the same  
symmetry algebra relations as the abbreviated non-invariant
operators , such as (12).
As a simple analogy of the above, one can think of the 
non-invariant symmetry
generators $\sim\oint{dz}\psi^m\psi^n$  inducing the truncated
global symmetries of (1) satisfying correct commutation
relations for Lorentz rotations in space-time. However,
the abbreviated non-invariant generators only act on $\psi$ and
not on $X$. To make them invariant, one has to add
extra terms proportional to $\int{\partial{X^{{\lbrack}m}}X^{n\rbrack}}$,
so that the invariant rotation generator satisfies the same symmetry algebra
but now acts on both $X$ and $\psi$.
We now turn to the question of symmetry algebras satisfied by the 
generators of the  type (5)-(7).
First of all, it is straightforward to check that, up to BRST-exact terms,
these operators all commmute with Poincare generators (2).
The geometrical meaning of the hidden symmetries
(3), (5)-(7) becomes clearer if  one considers
the vector analogues of these transformations given by
\eqn\grav{\eqalign{L^m=K\circ\oint{dz}{e^\phi}
(\lambda\partial^2{X^m}-2\partial\lambda\partial{X^m})\cr
L^{+}=K\circ\oint{dz}{e^\phi}(\lambda\partial^2\varphi
-2\partial\lambda\partial\varphi)\cr
L^{mn}=K\circ\oint{dz}\psi^m\psi^n\cr
L^{+m}=K\circ\oint{dz}\lambda\psi^m}}
Then, with some effort involving tedious picture-changing transformations ~{\spinselff}
one can  be shown  that the operators (11) 
realize the $AdS_{d+1}$ isometry algebra:
\eqn\grav{\eqalign{{\lbrack}L^m,L^n{\rbrack}=-L^{mn}\cr
{\lbrack}L^{+},L^{m}\rbrack=-L^{+m}\cr
{\lbrack}L^{m},L^{np}\rbrack=-\eta^{mn}L^p+\eta^{mp}L^n\cr
{\lbrack}L^{+},L^{mn}{\rbrack}=0\cr
{\lbrack}L^{+m},L^{np}{\rbrack}=-\eta^{mn}L^{+p}+\eta^{mp}L^{+n}\cr
{\lbrack}L^{mn},L^{pq}{\rbrack}=\eta^{mp}L^{nq}+...}}
It is worth mentioning that minus signs on the r.h.s. of the first
two commutators in (11), (12) appear in a rather nontrivial way, in a process
of cumbersome OPE calculations ~{\spinselff}.
The radial coordinate of the underlying $AdS_{d+1}$  
space related to the isometry algebra (12)
naturally coincides with the Liouville direction.
The operators of $AdS$ transvections $L^{m}$ and $L^{+}$ are the elements 
of $H_1$ and can also transformed into isomorphic $H_{-3}$ cohomology
by replacing 
$K\circ\oint{dz}\rightarrow{\oint{dz}},e^{\phi}\rightarrow{e^{-3\phi}}$.
The next step is to construct physical vertex operators based on isometry
generators (3), (5)-(7).
 Obviously the object of particular interest is spin 2 operator
in closed string sector, bilinear in transvection generators (11), (12),
with appropriate
momentum-dependent extra terms to ensure BRST-invariance.
As above, for our purposes, we shall limit ourselves to 
excitations polarized  and propagating along the
$AdS$ boundary (which in our case is simply
orthogonal to the Liouville
direction)
The construction in $H_{-3}\otimes{H_{-3}}$ cohomology leads to the following
expression for the operator:
\eqn\grav{\eqalign{
V_{s=2}=G_{mn}(p)\int{d^2z}e^{-3\phi-3{\bar\phi}}R^m{\bar{R}}^ne^{ipX}(z,{\bar{z}})\cr
R^m={\bar\lambda}\partial^2{X^m}-2\partial\lambda
\partial{X^m}
\cr+
ip^m({1\over2}\partial^2\lambda+{1\over{q}}
\partial\varphi\partial\lambda
-{1\over2}\lambda(\partial\varphi)^2+
(1+3q^2)\lambda(3\partial\psi_p\psi^p-{1\over{2q}}
\partial^2\varphi))\rbrace\cr
m=0,...,{d-1}
}}
where $G_{mn}$ is symmetric.
The operator on $H_1\otimes{H_1}$ is constructed likewise by replacing
$\int{d^2z}\rightarrow{K\bar{K}}\circ\int{d^2z}$ and
$-3\phi\rightarrow\phi,-3{\bar\phi}\rightarrow{\bar\phi}$.
Provided that $k^2=0$,0
it is straightforward to check its BRST-invariance
 with respect to the flat space BRST operator
\eqn\grav{\eqalign{Q_{brst}=\oint{dz}(cT-bc\partial{c}
-{1\over2}\gamma\psi_m\partial{X^m}-{1\over4}b\gamma^2)}}
as well as the linearized diffeomorphism invariance since the transformation
$G^{mn}(p)\rightarrow{G^{mn}}(p)+p^{(m}\epsilon^{n)}$  shifts holomorphic
and antiholomorphic factors of $V_{s=2}$ by terms
 BRST-exact in small Hilbert
space ~{\spinselff}.  In order to identify this symmetric massles spin 2 state 
with gravitational fluctuations, however, one needs to 
analyze the low-energy equations of motion for $G_{mn}$ which 
leading order is given by the Weyl constraints.
We will address this question in the next section.

\centerline{\bf 3. Flat vs AdS Gravitons: Weyl Invariance and Cohomology 
Structures }

As an instructive
 example, in this section we shall consider in detail the scale 
invariance constraints on the 
operator (13) of $H_{-3}\otimes{H_{-3}}$
 and compare them to those for the ordinary graviton in RNS theory (1).
To see the difference,
let us first recall the most elementary example - the graviton in bosonic
 string  theory given by
\eqn\lowen{V=G_{mn}\int{d^2z}\partial{X^m}{\bar\partial}{X^n}e^{ipX}}
The condition $\lbrack{Q},V\rbrack=0$  leads to constraints : 
$p^2G_{mn}(p)=p^mG_{mn}(p)=0$ 
related to linearized Ricci tensor contributions to the graviton's 
$\beta$-function.
The complete linearized contribution to the graviton's $\beta$-function, 
however, is
given by $\beta_{mn}=R_{mn}^{linearized}+2\partial_{m}\partial_{n}D$
(where $D$ is the space-time dilaton) with the last term 
particularly accounting for the
$\sim{e^{-2D}}$ factor in the low-energy effective action. 
This term in fact is $not$ produced
by any of the on-shell (BRST) constraints on the graviton vertex operator; 
to recover it, one
has to analyze the $off-shell$ constraints related to the Weyl invariance.
Namely, the generator of the Weyl transformations is given by,
the $T_{z{\bar{z}}}$ component of the stress-energy which is 
identically zero on-shell
in $d=2$ but nonzero  in $d=2+\epsilon$.
The leading order contribution to the $\beta-function$ of closed
string vertex operator $V$ is determined by the coefficient in front
of $\sim{{1\over{|z-w|^2}}}$ in the  midpoint OPE of
$T_{z{\bar{z}}}(z,{\bar{z}})$ and $V(w,{\bar{w}})$ leading to logarithmic
divergence in the integral 
$\sim\int{d^2z}\int{d^2w}T_{z{\bar{z}}}(z,{\bar{z}})V(w,{\bar{w}})$.
In bosonic string theory
one has 
\eqn\grav{\eqalign{
T_{z{\bar{z}}}\sim{-\partial{X_m}}{\bar\partial}{X^m}
+\partial\sigma{\bar\partial}\sigma+\partial{\bar\partial}(...)}}
skipping the full-derivative part proportional to $2d$ Laplacian
related to background charge, as it
leads to contact terms in the OPE with V, not contributing to its
$\beta$-function. 
Using (15) and (16) one easily calculates

\eqn\grav{\eqalign{
\int{d^2z}\int{d^2w}T_{z{\bar{z}}}(z,{\bar{z}})V(w,{\bar{w}})
\cr
\sim
G_{mn}(p)\int{d^2z}\int{d^2w}{{1\over{|z-w|^2}}}
e^{ipX}({{z+w}\over2},{{{\bar{z}}+{\bar{w}}}\over2})
\cr\times\lbrace
{p^2}\partial{X^m}{\bar\partial}{X^n}
-{1\over2}(p^mp_s\partial{X^s}{\bar\partial}{X^n}
+p^np_s\partial{X^m}{\bar\partial}{X^s})
\cr
+{1\over4}\eta^{mn}
p_sp_t\partial{X^s}{\bar\partial}{X^t}
\rbrace({{z+w}\over2},{{{\bar{z}}+{\bar{w}}}\over2})
\cr
\sim{ln\Lambda}{\lbrack}p^2G_{mn}(p)-{1\over2}
(p^sp_mG_{ns}(p)+p^sp_nG_{ms}(p))
\cr
+2p_mp_nD(p){\rbrack}
\int{d^2\zeta}\partial{X^m}{\bar\partial}{X^n}
e^{ipX}
(\zeta,{\bar\zeta})}}
where we introduced
$ln\Lambda=\int{{d^2\xi}\over{|\xi|^2}},\zeta=z+w$
and identified the dilaton with the trace of the space-time metric:
$D(p)\sim\eta^{st}G_{st}(p)$.

For conventional reasons and in  
order not to introduce too many letters in this paper
we adopt the same notation, $\Lambda$, for both the
worldsheet cutoff and the cosmological constant in space time.
However, we hope that the distinction between those will be very
clear to a reader from the context; in particular, in this paper
the cutoff shall always appears in terms of logs, while
all expressions in the cosmological  constants are either
linear or polynomial.

 The coefficient in front of the integral
thus determines the leading order contribution to the graviton's
$\beta$-function. The first three terms in this coefficient
simply give linearized Ricci tensor while the last one proportional
to the trace of the space-time metric determines the string coupling dependence.
All these terms contain two space-time derivatives and obviously no 
cosmological-type
contributions appear. The calculation analogous to (17) is of
 course similar in RNS superstring
theory, producing the similar answer. However,
in comparison with the bosonic string the RNS case
 also contains some instructive subtlety
which will be useful to observe for future calculations.
That is, consider graviton operator in RNS theory at canonical 
$(-1,-1)$-picture:
\eqn\lowen{V^{(-1,-1)}=G_{mn}(p)\int{d^2z}
e^{-\phi-{\bar\phi}}\psi^m{\bar\psi}^ne^{ipX}(z,{\bar{z}})}
The generator of Weyl transformations in $RNS$ theory is
\eqn\grav{\eqalign{T_{z\bar{z}}^{RNS}=T_{X}+T_{\psi}+T_{b-c}+T_{\beta-\gamma}+T_{Liouv}\cr
=-{1\over2}\partial{X_m}{\bar\partial}{X^m}-{1\over2}(\bar\partial\psi_m\psi^m+
\partial{\bar\psi}_m{\bar\psi}^m)\cr
+{1\over2}\partial\sigma\bar\partial\bar\sigma
-{1\over2}\partial\phi\bar\partial\bar\varphi
+{1\over2}\partial\chi\bar\partial\bar\chi+\partial\bar\partial(...)}}
The contribution of $T_X$ to the scale transformation of 
(18) is again easily computed to give
$\sim{p^2}G_{mn}(p)ln\Lambda{V^{(-1,-1)}}$, 
i.e. the gauge fixed linearized Ricci tensor
(with the gauge condition $p^m{G_{mn}}=0$ imposed by invariance under
transformations of ${V^{(-1,-1)}}$ by worldsheet superpartners of $T_X$, namely,
$G_{+{\bar{z}}}$ and $G_{-z}$.
To compute the contribution fom $T_\psi$, it is convenient to bosonize $\psi$
according to
\eqn\grav{\eqalign{
\psi_1\pm{i\psi_2}={e^{\pm{i}\varphi_1}}\cr
...\cr
\psi_{d-1}\pm{i\psi_d}=e^{\pm{i}\varphi_{{d}\over2}}
}}
(for simplicity we can assume the number $d$ of dimensions even, 
without loss of generality)
Then the stress-energy tensor for $\psi$ is
\eqn\grav{\eqalign{
T=-{1\over2}(\bar\partial\psi_m\psi^m+\partial\bar\psi_m\bar\psi^m)
=\sum_{i=1}^{{d\over2}}\partial\varphi_i\bar\partial\bar\varphi^i}}
Writing $\psi_1={1\over2}(e^{i\varphi_1}-e^{-i\varphi_1})$, it is easy to
compute the contribution of $T_\psi$ to the $\beta$-function:
\eqn\grav{\eqalign{
\int{d^2z}T_\psi(z,{\bar{z}})G_{mn}(p)\int{d^2w}e^{-\phi-{\bar\phi}}
\psi^m{\bar\psi}^ne^{ipX}(w,{\bar{w}})
\cr
={1\over2}ln\Lambda{G_{mn}}(p)\int{d^2\zeta}e^{-\phi-\bar\phi}
\psi^m{\bar\psi}^n{e^{ipX}}(\zeta,{\bar\zeta})
\equiv{1\over2}ln\Lambda
{V^{(-1,-1)}}}}
Note the scale transformation by $T_\psi$ contributes the term proportional to
$\sim{1\over2}G_{mn}$ with no derivatives, i.e. a ``cosmological'' type term.
The cosmological term is of course absent in 
the overall graviton's $\beta$-function as the contribution (22) 
is precisely cancelled by the scale transformation
of the ghost part of $V^{(-1,-1)}$ by 
$T_{\beta\gamma}=-{1\over2}|\partial\phi|^2+\partial\bar\partial(...)$:
\eqn\grav{\eqalign{
\int{d^2z}T_{\beta\gamma}(z,{\bar{z}})G_{mn}(p)\int{d^2w}e^{-\phi-{\bar\phi}}
\psi^m{\bar\psi}^ne^{ipX}(w,{\bar{w}})=
-{1\over2}ln\Lambda{V^{(-1,-1)}}}}
with the minus sign related to that of the $\phi$-ghost field in the 
trace of the stress-energy tensor.
So the absence of the cosmological
term in the graviton's $\beta$-function in RNS theory
is in fact the result of the smart cancellation between the Weyl transformations
of the matter and the ghost factors of the
graviton operator at $(-1,-1)$ canonical picture
(despite that the final answer - the absence of the overall
cosmological term may seem obvious)
The same result of course applies to the graviton operator (18) transformed
to any other ghost picture since it is straightforward to check that
both $\Gamma$ and $\Gamma^{-1}$ are Weyl-invariant, up to BRST-exact terms.
The absence of cosmological (or mass-like) terms in the Weyl transformation
laws is actually typical for any massless operators of $H_0$ or
$H_0\otimes{H_0}$; at nonzero pictures it is the  consequence of
the cancellation of Weyl transformations for the matter and the ghosts, as
was demonstrated above. This observation is of importance
since, as it will be shown below, this matter-ghost 
cancellation does $not$ occur for operators
of nonzero $H_n$'s, in particular, for the spin 2 operator (13)
in closed string theory and for massless 
operators for higher spin fields of Vasiliev type in open string sector.
Namely, we will show that
for the operator (13) the scale invariance 
constraints lead to cosmological term,
while for massless higher spin fields the similar constraints lead to 
emergence of AdS  geometry in the Fronsdal's operator in
the low-energy limit.
We start with analyzing the scale transformation of the
spin operator (13) by $T_X$. The canonical picture for the 
operator (13) is $(-3,-3)$.
To deduce the transformation law for the operator (13) it is sufficient
to consider the momentum-independent part 
$\sim{R_0^m}{\bar{R}}_0^n$
of the the matter factor
$\sim{R^m}{\bar{R}}^n$ in (13):
\eqn\grav{\eqalign{R^m=R_0^m+ik^m(...)\cr
R_0^m=\lambda\partial^2{X^m}-2\partial\lambda\partial{X^m}}}

and similarly for ${\bar{R}}^m$
Then the straightforward application of
$T_X$ to 
$$G_{mn}(p)\int{d^2w}e^{-3\phi-3{\bar\phi}}R_0^m{\bar{R}}_0^n{e^{ipX}}(w,{\bar{w}})$$
gives
\eqn\grav{\eqalign{
\int{d^2z}T_X(z,{\bar{z}})G_{mn}(p)\int{d^2w}e^{-3\phi-3{\bar\phi}}R_0^m{\bar{R}}_0^n{e^{ipX}}
(w,{\bar{w}})
\cr=
ln\Lambda\times{G_{mn}}(p)\int{d^2\zeta}
\lbrace{-{1\over2}p^2{e^{-3\phi-3\bar\phi}}
R_0^m{\bar{R}}_0^n{e^{ipX}}(\zeta,{\bar\zeta})
\rbrack}
\cr
-{i\over8}{p^m}\partial^2(e^{-3\phi}\lambda{e^{ipX}}(\zeta))
e^{-3{\bar\phi}}{\bar{R}}_0^n{e^{ipX}}({\bar{\zeta}})
\cr
+{i\over2}p^m\partial
\partial(e^{-3\phi}\partial\lambda{e^{ipX}}(\zeta))
e^{-3\bar\phi}{\bar{R}}_0^n{e^{ipX}}({\bar\zeta})
+(c.c.;m\leftrightarrow{n})\rbrack\rbrace
\cr
=ln\Lambda
{G_{mn}}(p)\int{d^2\zeta}
({-{1\over2}p^2\delta^{n}_q+{1\over2}p^mp_q})
{e^{-3\phi-3\bar\phi}}R_0^m{\bar{R}}_0^q{e^{ipX}}
+...}}
where we dropped BRST-exact terms and
 only kept terms contributing to the $G_{mn}$'s 
$\beta$-function, skipping those relevant to $\beta$-functions
of the space-time fields other than $G_{mn}$. In addition, for simplicity 
we skipped the dilaton-type contributions involving the trace of $G_{mn}$;
it is, however, straightforward to generalize the computation to include the dilaton,
accounting for the standard factor of $e^{-2D}$ in the effective action.
Comparing the transformation laws (17) and (25) one easily concludes that the 
contribution of $T_X$-transformation to the $G_{mn}$ $\beta$-function
results in the linearized Ricci tensor $R_{mn}^{linearized}$.
Next, consider the contributions from $T_\lambda=-{1\over2}
(\partial\bar\lambda\bar\lambda+\bar\partial\lambda\lambda)$
and $T_{\beta\gamma}$ to $\beta_{mn}$. The analysis is similar to the one
for the ordinary graviton operator (18)-(23), 
however, the crucial difference is that
this time there is no cancellation between transformations due to the worldsheet
matter (Liouville) fermion and the $\beta-\gamma$ ghost, observed above.
As previously, the transformation of (25) by $T_\lambda$ contributes
\eqn\grav{\eqalign{
G_{mn}(p)\int{d^2z}{T_\lambda}(z,{\bar{z}})\int{d^2w}
e^{-3\phi-3{\bar\phi}}R_0^m{\bar{R}}_0^ne^{ipX}(w,{\bar{w}})
\cr
={1\over2}ln\Lambda{G_{mn}}(p)\int{d^2\zeta}e^{-3\phi-3{\bar\phi}}R_0^m{\bar{R}}_0^ne^{ipX}(w,{\bar{w}})
}}
On the other hand, the transformation by $T_{\beta-\gamma}$ produces:
\eqn\grav{\eqalign{
G_{mn}(p)\int{d^2z}{T_{\beta-\gamma}}(z,{\bar{z}})\int{d^2w}
e^{-3\phi-3{\bar\phi}}R_0^m{\bar{R}}_0^ne^{ipX}(w,{\bar{w}})
\cr
=-{9\over2}ln\Lambda{G_{mn}}(p)\int{d^2\zeta}e^{-3\phi-3{\bar\phi}}R_0^m{\bar{R}}_0^ne^{ipX}(w,{\bar{w}})
}}
where we used the OPE $|\partial\phi|^2)z,{\bar{z}}e^{-3\phi-3{\bar\phi}}(w,{\bar{w}})
\sim{{9\over{|z-w|^2}}}e^{-3\phi-3{\bar\phi}}(w,{\bar{w}})$
Unlike the case of the ordinary graviton, the cosmological type
contributions from the scale transformations of the ghost and the matter part
of the operator (13) no longer cancel each other. 
As a result, the overall cosmological
term $\sim({9\over2}-{1\over2})G_{mn}$ appears 
in the $\beta$-function of (28) which leading order is now given by
\eqn\grav{\eqalign{\beta_{mn}=R_{mn}^{linearized}-8G_{mn}}}
(with the  extra factor of 2 related to the normalization of the Ricci tensor).
The appearance of the  cosmological term is thus closely related to the 
ghost cohomology structure of the operator (13), 
i.e. to the fact that the canonical
picture for this operator is $(-3,-3)$ while the standard $(-1,-1)$ 
picture representation
of the ``ordinary'' graviton does not exist for (13).
Collecting (25)-(27), this altogether allows us to identify
the space-time massless spin 2 $G_{mn}$ 
field emitted by $H_{-3}\otimes{H_{-3}}$ with
the gravitational fluctuations around the $AdS$ vacuum. In fact this 
is not a surprise since the operator (13) has been originally built
as a bilinear of the generators (11), (12) realizing transvections in $AdS$.
The next step is to generalize the above arguments to the vertex operators
for the massless higher spin fields (with $s\geq{3}$) which are also
the elements of nonzero cohomologies $H_{s-2}\sim{H_{-s}}$.
In analogy with the mechanism generating the cosmological term in (28), we
expect that the scale invariance analysis of these operators shall also lead
 to appearance of the mass-like terms in their $\beta$-functions
(although the operators by themselves are massless).  We shall attempt to show
that the ``mass-like'' terms are in fact related to the $AdS$ geometry couplings
of the higher spin fields, adding up to appropriate
$AdS$ Fronsdal operators in their
low-energy equations of motion in the leading order.

\centerline{\bf 4. Higher Spin Operators: Weyl Invariance and $\beta$-Functions}

In this section we extend the analysis of the previous sections to vertex 
operators
 describing massless higher spin excitations in open RNS string theory.
The space-time fields emitted by these operators  correspond to 
symmetric higher spin gauge
fields in Vasiliev's frame-like formalism.
 The main result of this section is that 
the leading order of the $\beta$-function for the higher spin operators
gives the low-energy equations of motion determined by Fronsdal operator
in the AdS space, despite the fact that the operators are initially 
defined around the flat background. As in the case of the AdS graviton 
considered in the 
previous section, the information about the AdS geometry is encrypted in the
ghost cohomology structure of the operators.
In the frame-like formalism {\fvf, \fvs, \vmaf, \vmas, \vmat, \vmafth, \svv,
\vcubic, \zhs, \skvnew} , a symmetric higher spin gauge field
of spin $s$ is described by collection of two-row fields
$\Omega^{s-1|t}\equiv\Omega_m^{a_1...a_{s-1}|b_1...b_t}(x)$ with $0\leq{t}\leq{s-1}$
and the rows of lengths $s-1$ and $t$. The only truly dynamical field of those
is $\Omega^{s-1|0}$ while the fields with $t\neq{0}$, 
called the extra fields, are related
to the dynamical one through generalized zero torsion constraints:
\eqn\lowen{\Omega^{s-1|t}\sim{D^{(t)}}\Omega^{s-1|0}}
where $D^{(t)}$ is certain order $t$ linear differential operator preserving
the symmetries of the appropriate Yang tableaux. There are altogether
$s-1$ constraints for the field of spin $s$.
As for the dynamical $\Omega^{s-1|0}$-field (symmetric in all the $a$-indices),
it splits into two diagrams with respect to the manifold $m$-index.
Assuming the appropriate pullbacks, the one-row symmetric diagram
describes the dynamics of the $metric-like$ symmetric Fronsdal's field
of spin $s$ while the two-row component of $\Omega^{s-1|0}$ can be removed
by appropriate gauge transformation.
In the language of string theory,  the higher spin $s$
 operators are the elements
 of $H_{s-2}\sim{H_{-s}}$.
The on-shell (Pauli-Fierz type) constraints on 
these space-time fields follow from 
the BRST-invariance constraints on the vertex operators, 
while the gauge transformations
correspond to shifting the vertex operators by 
BRST-exact terms (see ~{\seung} for detailed analysis).
The zero torsion constraints (29)  relating $\Omega^{s-1|t}$ 
gauge fields with different $t$
follow from the cohomology constraints on their vertex operators $V_{s-1|t}$, 
that is, by requiring
that all these vertex operators belong to the same cohomology 
$H_{s-2}\sim{H_{-s}}$
(there are, however, certain subtleties 
with this scheme arising at $t=s-1$ or $t=s-2$
which were discussed in ~{\seung} for the $s=3$ case)
The on-shell (Pauli-Fierz type) constraints on 
these space-time fields follow from 
the BRST-invariance constraints on the vertex operators, 
while the gauge transformations
correspond to shifting the vertex operators by BRST-exact terms 
(see ~{\seung} for detailed analysis).
Furthermore it turns out that the  vertex operators 
$V_{s-1|0}$  generating the $\Omega_m^{a_1...a_{s-1}}$
dynamical fields in space-time are only physical
 when $\Omega^{s-1|0}$ are fully symmetric
one-row fields (describing Fronsdal's metric-like tensors
 for symmetric fields of spin $s$)
while the operators for the two-row $(s-1,1)$-fields are typically the
 BRST commutators
in the small Hilbert space and therefore the space-time fields are pure gauge
 ~{\seung}.This altogether constitutes 
the dictionary between vertex operators in superstring theory 
(extended to higher ghost cohomologies).
Finally,
the zero torsion constraints (29)  relating 
$\Omega^{s-1|t}$ gauge fields with different $t$
follow from the cohomology constraints on their 
vertex operators $V_{s-1|t}$, that is, by requiring
that all these vertex operators belong to the same cohomology 
$H_{s-2}\sim{H_{-s}}$
(with some subtleties  at
 $t=s-1$ or $t=s-2$, mentioned above)
The zero torsion and cohomology constraints oinvolving the $t=s-1$
and $s-2$ cases  
are very interesting and deserve separate consideration, however,
 we shall not discuss them in this paper for the sake of brevity).

To understand the meaning of the cohomology constraints 
it is  useful to recall
first a much simpler
 example known from the conventional Ramond-Ramond sector
of closed superstring theory.
Namely, the relation between cohomology 
and zero torsion constraints can be thought of as a 
symmetric higher spin generalization of a more elementary and familiar
example  of standard Ramond-Ramond 
vertex operators in closed critical superstring theory.
It is well-known that the canonical picture representation for the Ramond-Ramond
operators is given by:
\eqn\grav{\eqalign{
V_{RR}^{(-{1\over2},-{1\over2})}={\not{F}}_{\alpha\beta}(p)
\int{d^2z}e^{-{\phi\over2}-{{\bar\phi}\over2}}
\Sigma^\alpha{\bar\Sigma}^\beta{e^{ipX}}(z,{\bar{z}})\cr
{\not{F}}_{\alpha\beta}^{(p)}\equiv\gamma^{m_1...m_p}_{\alpha\beta}F_{m_1...m_p}}}
where ${\not{F}}_{\alpha\beta}^{(p)}$ is the Ramond-Ramond $p$-form field
strength (contracted with $10d$ gamma-matrices)
Note that, since the operator (30) is the source of the field strength
( the derivative of the gauge potential), it does $not$ carry
RR charge (which instead is carried by a corresponding Dp-brane). The operator
(30) exists at all the pictures and is the element of
 $H^{(-{1\over2},-{1\over2})}$ cohomology
(which is the superpartner of $H^{(0,0)}$ 
consisting of all picture-independent
physical states).
 It is, however, possible to construct vertex operator which couples
to Ramond-Ramond gauge potential rather than field strength.
The canonical picture for such an operator is $(-{3\over2},-{1\over2})$
(or equivalently  $(-{1\over2},-{3\over2})$
with the explicit expression given by

\eqn\grav{\eqalign{
U_{RR}^{(-{1\over2},-{3\over2})}={\not{A}}_{\alpha\beta}(p-1)
\int{d^2z}e^{-{\phi\over2}-{{3\bar\phi}\over2}}
\Sigma^\alpha{\bar\Sigma}^\beta{e^{ipX}}(z,{\bar{z}})\cr
{\not{A}}_{\alpha\beta}^{(p-1)}\equiv\gamma^{m_1...m_p}_{\alpha\beta}A_{m_1...m_{p-1}}}}
where generically,  ${\not{A}}$ is arbitrary.
The $U$-operator (31) is generally not the picture-changed version
of the $V$-operator (30)  nor it is the element of
$H^{(-{1\over2},-{1\over2})}$ for general ${\not{A}}$. 
To relate $U_{RR}^{(-{1\over2},-{3\over2})}$ to
$V_{RR}^{(-{1\over2},-{1\over2})}$ of (30)
 by the picture-changing:
\eqn\lowen{V_{RR}^{(-{1\over2},-{1\over2})}=:{\Gamma}U_{RR}^{(-{3\over2},-{1\over2})}:}
one has to impose the constraint ${\not{F}}=d{\not{A}}$
that ensures that $U$ is the 
physical operator of $H^{(-{1\over2},-{1\over2})}$. Thus the cohomology
constraint in $U$ leads to the standard relation between the gauge potential
and the field strength. Similarly,  the generalized 
$H_{s-2}\sim{H_{-s}}$-cohomology constraints
on higher spin operators $V_{s-1|t}$ for $\Omega^{s-1|t}$ space-time fields
lead to generalized zero torsion constraints (29). Note that for 
$0\leq{t}\leq{s-3}$
the canonical pictures for $V_{s|t}$ are ${2s-t-5}\sim{t+3-2s}$
with the cohomology constraints $V_{s|t}{\in}H_{s-2}\sim{H_{-s}}$
inducing the chain (29) of zero torsion relations.

With are now prepared to analalyze the scale invariance constraints
for open string vertex operators describing the Vasiliev
 type higher spin fields
in space-time. It turns out that
for  massless fields of spin $s$ the canonical picture representation
is especially simple for the field with $t=s-3$, that is, for
$\Omega^{s-1|s-3}$. 
The explicit vertex operator expression for this field is given by:

\eqn\grav{\eqalign{
V_{s-1|s-3}=
\Omega_m^{a_1...a_{s-1}|b_1...b_{s-3}}(p)
\int{d^2z}e^{-s\phi-s{\bar\phi}}
\cr\times
\psi^m\partial\psi_{b_1}\partial^2\psi_{b_2}
...\partial^{s-3}\psi_{b_{s-3}}\partial{X_{a_1}}...\partial{X_{s-1}}e^{ipX}
\cr
\sim
\Omega_m^{a_1...a_{s-1}|b_1...b_{s-3}}(p)K\circ
\int{d^2z}e^{(s-2)\phi-(s-2){\bar\phi}}
\cr\times
\psi^m\partial\psi_{b_1}\partial^2\psi_{b_2}
...\partial^{s-3}\psi_{b_{s-3}}\partial{X_{a_1}}...\partial{X_{s-1}}e^{ipX}}}

For s=3, this immediately gives the operator for the Fronsdal field considered
in ~{\spinself, \seung}.
The on-shell conditions on $\Omega^{s-1|s-3}$ to ensure 
the BRST-invariance of (33) are not difficult to 
obtain using the BRST charge (14).
The commutation with the $T_{X}$ component 
of the stress-energy part of $Q_{brst}$
leads to the tracelessness of $\Omega$ in the $a$-indices,
that is, $\Omega_{ma}^{aa_1...a_{s-3}|b_1...b_{s-3}}=0$ which is the
 well-known constraint
on frame-like fields and to the second Pauli-Fierz constraint of
transversality: $p_a\Omega_m^{aa_1...a_{s-2}|b_1...b_{s-3}}(p)=0$
The commutation with $T_\psi$ part of $Q_{brst}$, given by
$-{1\over2}\oint{dz}c\partial\psi_p\psi^m$, 
requires the symmetry of $\Omega$ in the
$b$-indices, as it is easy to see from the OPE between $T_\psi$
and $S_{mb_1...b_{s-3}}=\psi_m\partial\psi_{b_1}...\partial^{s-3}\psi_{b_{s-3}}$ 
- the latter is the primary
field of dimension $h_\psi={1\over2}(s-2)^2$
only if $S$ is symmetric and traceless in all indices.
While the symmetry in the $b$-indices is another standard familiar 
constraint in the frame-like formalism,
the symmetry and tracelessness of $m$ with respect to the $b$-indices
is an extra condition on $\Omega$ that can be obtained partial fixing of 
the gauge symmetries of $\Omega$.  Given the above conditions are 
fulfilled, the
commutation with the supercurrent part of $Q_{brst}$ produces no new 
constraints, however, there is one more condition coming from the
$H_{-s}$-cohomology constraint on $V_{s-1|s-3}$, that is,
\eqn\lowen{:\Gamma{V_{s-1|s-3}}:=0}
This constraint further requires the vanishing of the mixed trace
over any pair of $(a,b)$=indices:
$\eta_{ab}\Omega_m^{aa_1...a_{s-2}|bb_1...b_{s-4}}=0$.
Fortunately the gauge symmetry of $\Omega$ is more than
powerful enough to absorb this extra constraint as well.
Finally, we are left to consider the BRST nontriviality
conditions on (33).
First of all, the nontriviality constraint: 
$V_{s-1|s-3}\neq{\lbrace}Q_{brst},W_{s-1|s-3}\rbrace$ 
where $W$ is some operator in small Hilbert space
requires either
\eqn\grav{\eqalign{\eta^m_a\Omega_m^{aa_1...a_{s-2}|b_1...b_{s-3}}\neq{0}}}
or
\eqn\grav{\eqalign{p^m\Omega_m^{a_1...a_{s-1}|b_1...b_{s-3}}\neq{0}}}
since otherwise, generically, there exist operators
\eqn\grav{\eqalign{
W_{s-1|s-3}\sim\Omega_m^{a_1...a_{s-1}|b_1...b_{s-3}}\sum_{k=0}^{s-1}\oint{dz}
{e^{\chi-(s-1)\phi}}\partial\chi
\cr\times
\partial\psi_{b_1}...\partial^{s-3}\psi_{b_{s-3}}
\partial{X_{a_1}}...\partial{X_{s-1}}
\partial^{s-1-k}X_mG^{(k)}(\phi,\chi)
e^{ipX}}}
commuting with the stress tensor part of $Q_{brst}$ while, at the same time,
the commutators of the supercurrent
part of $Q_{brst}$ with $W_{s-1|s-3}$ are proportional to
$V_{s-1|s-3}$:
${\lbrace}Q_{brst},W_{s-1|s-3}\rbrace=\alpha_sV_{s-1|s-3}$
where $\alpha_s$ are some numbers (generically, nonzero)
and $G^{(k)}(\phi,\chi)$ are polynomials in derivatives
of $\phi$ and $\chi$ of conformal dimension $k$ ( generically, 
inhomogeneous in degree
and quite cumbersome)
such that
$$e^{\chi-(s+1)\phi}\partial^{s-1-k}X_mG^{(k)}(\phi,\chi)\partial\chi$$
 is a primary field
(this is a rather stringent constraint which, nevertheless, typically
has nontrivial solutions for generic $s$;
e.g.  see ~{\spinself} for some concrete examples).
For this reason, unless one of the nontriviality conditions  (35) or (36) 
holds, 
the operators $V_{s-1|s-3}$ are BRST-exact in small Hilbert space;
however, if either (35) or (36) are satisfied, 
the $W$-operators do not commute with the
stress-energy tensor part of $Q_{brst}$ and therefore
 their overall commutators with
$Q_{brst}$ no longer produce $V_{s-1|s-3}$ with the latter 
now being in BRST cohomology
and physical. However, it is easy to see  that out of 2 possible nontriviality 
conditions (35), (36)
it is the second one (36) that must be chosen since the 
first one clearly violates
the $H_{-s}$-cohomology condition (34).
This immediately entails the gauge transformations for the $\Omega$-field:
\eqn\grav{\eqalign{
\Omega_m^{a_1...a_{s-1}|b_1...b_{s-3}}\rightarrow
\Omega_m^{a_1...a_{s-1}|b_1...b_{s-3}}+p_m\Lambda^{a_1...a_{s-1}|b_1...b_{s-3}}}}
that shift $V_{s-1|s-3}$ by BRST-trivial terms irrelevant for amplitudes and
lead to well-known vast and powerful gauge symmetries possessed 
by the higher spin fields.
Note that, although all the above analysis has been 
performed  for the operators
at negative cohomologies (which are simpler from the technical point of view),
all the above results directly apply to the corresponding operators at
isomorphic positive $H_{s-2}$-cohomologies since the explicit isomorphism between
negative and positive cohomologies is BRST-invariant ~{\spinself}.
 
To complete our analysis of BRST on-shell constraints on the higher 
spin operators of $H_{s-2}\sim{H_{-s}}$
we shall comment on the only remaining possible source of BRST-triviality
for $V_{s-1|s-3}$ coming from operators proportional to the ghost factor
${\sim}e^{2\chi-{(s+2)}\phi}$.
All the hypothetical operators in the small Hilbert space with 
such a property are given by:
\eqn\grav{\eqalign{U_{s-1|s-3}=
\Omega_m^{a_1...a_{s-1}|b_1...b_{s-3}}
\oint{dz}c\partial\xi\partial^2\xi{e^{-(s+2)\phi}}R^{(2s-2)}(\phi,\chi,\sigma)
\cr\times
\psi^m\partial\psi_{b_1}...\partial^{s-3}\psi_{b_{s-3}}
\partial{X_{a_1}}...\partial{X_{s-1}}e^{ipX}}}
where $R^{(2s-2)}$ is the conformal dimension
$2s-2$ polynomial in derivatives of $\phi$,$\chi$ and $\sigma$
(again, homogeneous in conformal weight but not in degree).
Indeed, the commutator of the matter supercurrent part of $Q_{brst}$, 
given by $-{1\over2}\oint{dw}\gamma\psi_m\partial{X^m}$ with 
$U_{s-1|s-3}$ is zero since the leading order of the OPE between
$\gamma\psi_m\partial{X^m}(w)$ and the integrand of $U_{s-1|s-3}$ at a point $z$ is
nonsingular, that is, proportional to $(z-w)^0$, as is easy to check.
At the same time , the commutator of ${U_{s-1|s-3}}$ with 
 the ghost supercurrent part of $Q_{brst}$, given by
$-{1\over4}b\gamma^2$, is nonzero and is proportional to $V_{s-1|s-3}$:
\eqn\lowen{{\lbrace}{Q_{brst}},{U_{s-1|s-3}}\rbrace=\lambda_s{V_{s-3}}}

(where $\lambda_s$ are certain numbers) provided that
the coefficient $\sigma_{2s-2}$ in front of the leading OPE order
of $R^{(2s-2)}$ and $b\gamma^2$ is nonzero:
\eqn\grav{\eqalign{
R^{(2s-2)}(z):b\gamma^2:(w)\sim{{{\sigma_{2s-2}}b\gamma^2(w)}\over{(z-w)^{2s-2}}}
+O(z-w)^{2s-3}
\cr
\sigma_{2s-2}\neq{0}}}
Then, provided that the conditions
\eqn\lowen{\lambda_s\neq{0}}
and
\eqn\lowen{\sigma_{2s-2}\neq{0}}
are both satisfied,
 the operator $V_{s-1|s-3}$ could be trivial only if the 
stress-tensor part of $Q_{brst}$ commuted with $U_{s-1|s-3}$  which is only
possible if
(given the on-shell conditions on $\Omega$ described above)
\eqn\lowen{
G_s(z)=:c\partial\xi\partial^2\xi{e^{-(s+2)\phi}}R^{(2s-2)}(\phi,\chi,\sigma):(z)}
is a primary field. That is,
the OPE of $G_s$ 
with the full ghost stress-energy tensor:
\eqn\grav{\eqalign{
T_{gh}={1\over2}(\partial\sigma)^2
+{1\over2}(\partial\chi)^2-{1\over2}(\partial\phi)^2
+{3\over2}\partial^2\sigma+{1\over2}\partial^2\chi-\partial^2\phi}}
is
generically given by
\eqn\grav{\eqalign{T_{gh}(z)G_s(w)=
\sum_{k=0}^{2s-1}{{y_{k}Y^{(-{{s^2}\over2}-s+k)}(w)}\over{(z-w)^{2s+2-k}}}
+{{(s-{1\over2}s^2){G_s}(w)}\over{(z-w)^2}}
+{{{\partial}{G_s}(w)}\over{(z-w)}}+O(z-w)^0}}
where $y_k$ are numbers and
$Y^{(-{{s^2}\over2}-s+k)}$ are operators of 
conformal dimensions $-{{s^2}\over2}-s+k$.
So   the $V_{s-1|s-3}$ operators trivial
only if the constraints
\eqn\grav{\eqalign{y_k=0\cr
k=0,...,2s-1}}
are fulfilled  simultaneously with the conditions (42), (43).
Clearly, for $s$ large enough the constraints (42), (43), (47) 
are altogether too restrictive,
leaving no room for any  possible choice of 
$R^{(2s-2)}(\phi,\chi,\sigma)$, so the operators
are nontrivial (of course, provided that (36) holds as well).
 To see this note that, for any large $s$  and given $k$ 
in the sum (46)  the number of independent
operators $Y^{(-{{s^2}\over2}-s+k)}$ is of the order of 
${\sim}{{d}\over{dk}}({{e^{{a}{\sqrt{2s-k}}}}\over{\sqrt{2s-k}}})$
where $a$ is certain constant,
since the number of conformal weight $n$ polynomials is of the order of 
the number of partitions of $n$ which, in turn, is given
by Hardy-Ramanujan  asymptotic formula for large $n$.
 Summing over $k$, it is clear that the number
of constraints (47) on $G^{(2s-2)}$ is asymptotically of the order of
${{e^{{a}{\sqrt{s}}}}\over{\sqrt{s}}}$ 
while the number of independent terms in
$R^{(2s-2)}$ is of the order of 
${{e^{{a}{\sqrt{s}}}}\over{{s}}}$, so the number of constraints (47) 
exceeds the number of possible
operators $U_{s-1|s-3}$ by  the factor of the order of ${\sqrt{s}}$.
Therefore all the operators (33) with large spin values  are BRST-nontrivial,
provided that (36) is satisfied.
 For the lower
values of $s$, however, the constraints (42), (43), (47) 
have to be analyzed separately. For $s=3,4$
 it can be shown that the constraints (43), (47) lead to polynomials satisfying
$\lambda_s=0$, so the appropriate higher spin operators are physical.
For $5\leq{10}$ direct numerical analysis shows the incompatibility
of the conditions (42),(43), (47) with the number of constraints 
exceeding the number of operators of the type (39) posing a
potential threat of BRST-triviality, showing that operators 
with spins greater than 4 are physical
as well. 

With the on-shell BRST conditions  pointed  out, 
the next step is to analyze
the scale invariance (off-shell) constraints on the operators (33).
It is instructive to start with the $s=3$ case since for  $s=3$
$\Omega_{s-1|s-3}$ is precisely the Fronsdal's field.
Similarly to the closed string case, the Weyl transformation of $V_{s-1|s-3}$
is determined by the OPE coefficient in front of $\sim{|z-\tau|^{-2}}$
term in the operator product
$lim_{z,{\bar{z}}\rightarrow\tau}T^{z{\bar{z}}}(z,{\bar{z}})V_{s-1|s-3}(\tau)$
where $\tau$ is on the worldsheet boundary and, as previously, the
$\epsilon$-expansion setup is assumed, so $T^{z{\bar{z}}}\neq{0}$.
Starting
 from the transformation by $T_{X}=-{1\over2}|\partial{{\vec{X}}}|^2$, we have
\eqn\grav{\eqalign{\int{d^2z}T_{X}^{z{\bar{z}}}(z,{\bar{z}})
\Omega_m^{a_1a_2}(p)\oint{d\tau}e^{-3\phi}\psi^m\partial
{X_{a_1}}\partial{X_{a_2}}e^{ipX}(\tau)
\cr\sim
ln\Lambda\times
\oint{d\tau}e^{-3\phi}\psi^m\partial
{X_{a_1}}\partial{X_{a_2}}e^{ipX}(\tau){\lbrack}
-p^2\Omega_m^{a_1a_2}(p)+2p_tp^{(a_1}\Omega^{a_2)t}_m-
p^{a_1}p^{a_2}\Omega^{\prime}_m\rbrack}}
where we introduced $\Omega^{\prime}_m\equiv\eta_{a_1a_2}\Omega_m^{a_1a_2}$
(similarly, using the
Fronsdal's notations, the ``prime'' will stand for contraction
of a pair of fiber 0indices for any other higher spin field below)
This gives the part of the leading order contribution to the spin 3 
$\beta$-function proportional to the Fronsdal's
operator in flat space. The analysis of the
 contributions by $T_\psi^{z{\bar{z}}}$ and by 
$T_{\beta-\gamma}^{z{\bar{z}}}$ is analogous 
to the one performed in the previous section 
for the AdS graviton operator (13) and the result is
\eqn\grav{\eqalign{\int{d^2z}(T_{\psi}^(z,{\bar{z}})+T_{\beta-\gamma}^(z,{\bar{z}}))
\Omega_m^{a_1a_2}(p)\oint{d\tau}e^{-3\phi}\psi^m\partial
{X_{a_1}}\partial{X_{a_2}}e^{ipX}(\tau)
\cr\sim
-8{ln\Lambda}\Omega_m^{a_1a_2}\times
\oint{d\tau}e^{-3\phi}\psi^m\partial
{X_{a_1}}\partial{X_{a_2}}e^{ipX}(\tau)}}
where the coefficient in front of $\Omega$ 
ensures that the overall normalization of
(49) is consistent with that of (48).
As in the case of the cosmological term appearing 
in the graviton's $\beta$-function (28),
the appearance of the mass-like term in the spin 3 $\beta$-function 
(48), (49) is due to the 
non-cancellation of the corresponding terms  in 
the Weyl transformation laws for the matter
and for the ghost parts, which in turn 
is the consequence of the $H_{-3}\sim{H_1}$-cohomology
coupling of the spin 3 operator. 
The term (49) in the $\beta$-function is $not$, however,
a mass term.
Namely, combined together, the  contributions (48), (49) 
give the low-energy equations of motion for
massless spin $3$ field, corresponding to the special case  of the Fronsdal's
operator in the $AdS$ space acting on spin 3 field that is polarized along
the $AdS$ boundary and is propagating parallel to the boundary.
The correspondence between (49) and the mass-like term in the Fronsdal's
operator in $AdS_{d+1}$ is exact for $d=4$; 
to make the correspondence precise for $d\neq{4}$ requires
some modification of the operators of the type (33) (see the discussion
below for general spin case).

The next step is to generalize this simple calculation 
to the general spin value
and to calculate the $\beta$-functions of the frame-like fields (33). 
The vertex operators (33) do not generate Fronsdal's fields for $s\geq{4}$
(but rather the derivatives of the Fronsdal's fields), 
and explicit expressions for
$V_{s-1|t}$-operators for $0\leq{t}\leq{s-4}$, following from cohomology 
constraints are generally quite complicated.
For example, the manifest form of operators for $\Omega_{s-1|s-4}$-fields
at canonical $(-s-1)$-picture is given by
\eqn\grav{\eqalign{
V_{s-1|s-4}=\Omega_m^{a_1...a_{s-1}|b_1...b_{s-4}}(p)\oint{dz}
e^{-(s+1)\phi}\psi_m\partial\psi^{b_1}...\partial^{s-4}\psi^{b_{s-4}}
\cr
{\times}\sum_{k=0}^{2s-3}T^{(2s-3-k)}(\phi)
\lbrack\sum_{j=1}^{k-1}a_j\partial^jX_q\partial^{k-j}X^q+b_j
\partial^{j-1}\psi_q\partial^{k-j}\psi^q\rbrack}}
where  $a_j$ and $b_j$ are certain coefficients and
$T^{(2s-3-k)}(\phi)$ are again certain conformal dimension
$2s-3-k$ inhomogeneous polynomials in the derivatives of  $\phi$.
The coefficients and the polynomial structures must be chosen to ensure that
the integrand of (50) is primary 
field of dimension 1 and the picture-changing transformation
of the operator (50) is nonzero, producing an operator
 at picture $-s$ and at cohomology
$H_{-s}$,
 so that the hohomology condition
on (50) produces the zero torsion-like condition relating the frame-like fields 
in Vasiliev's formalism:
\eqn\grav{\eqalign{:\Gamma{V_{s-1|s-4}}:=
{V_{s-1|s-3}}+{\lbrace}Q_{brst},...\rbrace\cr
\Omega^{s-1|s-3}(p)\sim{p}\Omega^{s-1|s-4}(p)}}
so that the transformation of
${V_{s-1|s-4}}$ produces the vertex operator proportional to ${V_{s-1|s-3}}$
with the space-time field 
$\Omega^{s-1|s-3}(p)\sim{p}\Omega^{s-1|s-4}(p)$ given by certain first order 
differential operator acting on $\Omega^{s-1|s-4}(p)$. The explicit structure 
of this operator  (giving one of the zero curvature constraints)
is  determined by the details of the picture-changing;
for example one of the contributions to (50) from
the picture transformation of the  $k=0$ term in (50) results 
from the OPE contributions:
\eqn\grav{\eqalign{e^\phi(z){e^{-(s+1)\phi}}(w)
\sim{(z-w)^{s+1}}{e^{-s\phi}}({{z+w}\over2})+...\cr
\partial{X^q}(z){e^{ipX}}(w)\sim(z-w)^{-1}
\times{(-ip^q)}e^{ipX}({{z+w}\over2})+...\cr
{e^{\phi}}(z)T^{(2s-3)}(\phi)(w)\sim(z-w)^{3-2s}{e^{\phi}}({{z+w}\over2})+...}}
so the leading OPE order of the product of the  picture-changing operator
$\Gamma\sim{-{1\over2}{e^\phi}\psi_q\partial{X^q}}+...$ with $V_{s-1|s-4}$ is
 ${\sim}(z-w)^{3-s}$,
so to obtain the normally ordered contribution, 
relevant to the picture-changing
transformation (51), one has to expand the remaining field
$\psi_q(z)$ of $\Gamma$ up to the order of $s-3$ around the midpoint 
${{z+w}\over2}$
which altogether produces the result proportional to $V_{s-1|s-3}$, 
with the space-time 
field  proportional to the space-time derivative of $\Omega^{s-1|s-4}$
(as it is clear from the second OPE in (52)). 
There are of course many other terms
in the OPE between $\Gamma$ and $V_{s-1|s-4}$ but, provided that all the coefficients
and the polynomial structures in (50) are chosen correctly,
 they all give the result
proportional to $V_{s-1|s-3}$, up to
 BRST-exact  terms and with the zero torsion condition:
$$\Omega^{s-1|s-3}\sim{\partial}\Omega^{s-1|s-4}$$ 
controlled by the picture-changing
procedure. The explicit expressions for the operators with
$t=s-5,s-6,...$ and, ultimately, for $t=0$ (Fronsdal's field) are 
increasingly complicated
for general $s$. 
However, in order to deduce the Weyl invariance constraints on 
massless vertex operators for Fronsdal's fields  of spin $s$, 
we don't actually need to know
the explicit expressions for $V_{s-1|0}$. 
The key point here is the mutual independence of
the Weyl transformations and the cohomology constraints 
on the vertex operators. That is, the
cohomology constraints relate
the Fronsdal's operator at canonical $3-2s$-picture 
and  the operator for  the $\Omega^{s-1|s-3}$ extra-field
through
\eqn\grav{\eqalign{\Omega_m^{a_1...a_{s-1}|b_1...b_{s-3}}(p)\oint{dz}
e^{-s\phi}\psi^m\partial\psi_{b_1}...\partial^{s-3}\psi_{b_{s-3}}
\partial{X_{a_1}}...\partial{X_{a_{s-1}}}e^{ipX}
\cr
=\Omega_m^{a_1...a_{s-1}}:\Gamma^{s-3}:\oint{dz}U^{m(-2s+3)}_{a_1...a_{s-1}}(p)}}
where $U$ is the indegrand of the vertex operator for the Fronsdal's field.
Since $\Gamma$ is BRST and Weyl-invariant, 
the relation (53) allows to deduce the low-energy
equations of motion for the Fronsdal's fields by studying 
the Weyl transformations of the operators
(33) which are much simpler.
The transformations of $V_{s-1|s-3}$
by $T_X^{z{\bar{z}}}$ and $T_{\beta-\gamma}^{z\bar{z}}$ are 
computed similarly to the spin 3 case considered above.
One easily finds
\eqn\grav{\eqalign{
\int{d^2z}T_X^{z{\bar{z}}}(z,{\bar{z}})
\Omega_m^{a_1...a_{s-1}|b_1...b_{s-3}}(p)\oint{d\tau}
e^{-s\phi}\psi^m\partial\psi_{b_1}...\partial^{s-3}\psi_{b_{s-3}}
\partial{X_{a_1}}...\partial{X_{a_{s-1}}}e^{ipX}
\cr
\sim{ln}\Lambda\oint{d\tau}
e^{-s\phi}\psi^m\partial\psi_{b_1}...\partial^{s-3}\psi_{b_{s-3}}
\partial{X_{a_1}}...\partial{X_{a_{s-1}}}e^{ipX}
\cr\times
\lbrack
-p^2\Omega_m^{a_1...a_{s-1}|b_1...b_{s-3}}(p)+
p_t{\Sigma_1}(a_1|a_2...a_{s-1})p^{a_1}\Omega_m^{a_2...a_{s-1}t|b_1...b_{s-3}}
\cr
-{1\over2}\Sigma_2(a_{s-2},a_{s-1}|a_1,...,a_{s-3})p^{a_{s-1}}p^{a_{s-2}}
(\Omega_m^\prime)^{a_1...a_{s-3}|b_1...b_{s-3}}}}
and
\eqn\grav{\eqalign{
\int{d^2z}T_{\beta-\gamma}^{z{\bar{z}}}(z,{\bar{z}})
 \Omega_m^{a_1...a_{s-1}|b_1...b_{s-3}}(p)\oint{d\tau}
e^{-s\phi}\psi^m\partial\psi_{b_1}...\partial^{s-3}\psi_{b_{s-3}}
\partial{X_{a_1}}...\partial{X_{a_{s-1}}}e^{ipX}
\cr\sim{-}s^2\Omega_m^{a_1...a_{s-1}|b_1...b_{s-3}}(p){ln}\Lambda
\oint{dz}
e^{-s\phi}\psi^m\partial\psi_{b_1}...\partial^{s-3}\psi_{b_{s-3}}
\partial{X_{a_1}}...\partial{X_{a_{s-1}}}e^{ipX}}}
Here
 $\Sigma_1(b|a_1...a_n)$ and $\Sigma_2(b_1,b_2|a_1....a_n)$ 
are the Fronsdal's symmetrization operations
~{\fronsdalsit}, acting on free indices, e.g. 
$\Sigma_p(a_1,...,a_p|b_1,...,b_s)T^{aa_1...a_p}H_a^{b_1...b_s}$  where $H$ is symmetric, symmetrizes
over $a_1,...a_p;b_1,...,b_s$.

To compute the Weyl transform of the $\psi$-part, it is again helpful
to use the bosonization relations (20), (21). 
Since the bosonized $\varphi_i$ fields 
carry no background charges (as it is clear from the stress-energy tensor (21)),
the coefficient in front of the $|z-\tau|^2$ term in the OPE
of $T_\psi^{z\bar{z}}(z,{\bar{z}})$ and $V_{s-1|s-3}(\tau)$
coincides with  the conformal dimension
of the $\psi$-factor:
$\psi^m\partial\psi_{b_1}...\partial^{s-3}\psi_{b_{s-3}}$ which is
equal to ${1\over2}(s-2)^2$, so
\eqn\grav{\eqalign{
\int{d^2z}T_{\psi}^{z{\bar{z}}}(z,{\bar{z}})
 \Omega_m^{a_1...a_{s-1}|b_1...b_{s-3}}(p)\oint{d\tau}
e^{-s\phi}\psi^m\partial\psi_{b_1}...\partial^{s-3}\psi_{b_{s-3}}
\partial{X_{a_1}}...\partial{X_{a_{s-1}}}e^{ipX}
\cr
\sim{(s-2)^2\Omega_m^{a_1...a_{s-1}|b_1...b_{s-3}}(p){ln}\Lambda
\oint{dz}
e^{-s\phi}\psi^m\partial\psi_{b_1}...\partial^{s-3}\psi_{b_{s-3}}
\partial{X_{a_1}}...\partial{X_{a_{s-1}}}e^{ipX}+...}}}
(again, with no factor of ${1\over2}$
due to the normalization chosen for the kinetic term).
The last identity is true as long as the $\psi$-factor is a primary
field, i.e. the appropriate on-shell conditions are imposed on
$\Omega$. It is not difficult to see, however, that the
contributions due to the off-shell part are generally
proportional to space-time derivatives of $\Omega$ and
its traces, multiplied by higher spin operators that are not of
the form (33),  so these contributions are irrelevant
for $\beta$-functions of the higher spin fields of Vasiliev's type
(instead, they contribute to the low-energy equations of motion
of more complicated higher spin fields, such as those with mixed symmetries;
so these contributions may become important in various generalizations
of the Vasiliev's theory).
Collecting (54)-(56) and using the cohomology constraint (53), we deduce that 
the leading order $\beta$-function for the massless Fronsdal's fields of 
spin $s$ is  
\eqn\grav{\eqalign{\beta_m^{a_1...a_{s-1}}=
-p^2\Omega_m^{a_1...a_{s-1}}(p)
+
{\Sigma_1}(a_1|a_2,...a_{s-1})p_tp^{a_1}\Omega_m^{a_2...a_{s-1}t}
\cr
-{1\over2}\Sigma_2(a_{s-2},a_{s-1}|a_1,...,a_{s-3})p^{a_{s-1}}p^{a_{s-2}}
(\Omega_m^\prime)^{a_1...a_{s-3}}
-4(s-1)\Omega_m^{a_1...a_{s-1}}}}

The appearance of the mass-like terms is  related to the emergence
of the curved geometry already observed in (28).
Namely, vanishing of the $\beta$-function (57) gives, in the leading order,
the low-energy  effective equations of motion on $\Omega$
given by
\eqn\lowen{{\hat{F}}_{AdS}\Omega=0}
where 
${\hat{F}}_{AdS}$ is the Fronsdal's operator in $AdS_{d+1}$ space
,(exactly for $d=4$ and with some modifications in other dimensions)
which action is restricted
 on higher spin fields $\Omega$ polarized along the $AdS$ boundary.
Indeed, 
the explicit expression for the Fronsdal's operator in $AdS_{d+1}$ 
  ~{\fronsdalsit}, acting on symmetric spin $s$ 
fields polarized along the
boundary is:
\eqn\grav{\eqalign{({\hat{F}}_{AdS}\Omega)^{a_1...a_s}
=\nabla_A\nabla^A\Omega^{a_1...a_s}-\Sigma_1
(a_1|a_2...a_s)\nabla_t\nabla^{(a_1}
\Omega^{a_2...a_st)}
\cr
+{1\over2}\Sigma_2(a_1,a_2|a_3,...,a_s)
\nabla^{a_1}\nabla^{a_2}(\Omega^\prime)^{a_3...a_s}
-m_\Omega^2\Omega^{a_1...a_s}+2\Sigma_2\Lambda{g^{a_1a_2}}(\Omega^\prime)^{a_3...a_s}
\cr
m^2_\Omega=-\Lambda(s-1)(s+d-3)
}}
where $A=(a,\alpha)$ is the $AdS_{d+1}$ space-time index
(with the latin indices being along the boundary and $\alpha$ 
being the radial direction) 

The cosmological constant in our units is fixed 
 $\Lambda=-4$, to make it consistent with the Weyl transform 
of the $AdS$ graviton operator (13) 
In what follows we shall ignore
the last term in this operator since, in the string theory context,
it is related  to the higher-order (cubic) contributions to the
$\beta$-function, which are
beyond the leading order Weyl invariance constraints.
For the remaining part, consider the box ($\nabla^2$) of $\Omega$
first. It is convenient to use the Poincare coordinates
for $AdS$:
\eqn\lowen{ds^2={{R^2}\over{y^2}}(dy^2+dx_adx^a)}.
With the Christoffel's symbols:
\eqn\grav{\eqalign{
\Gamma^y_{a_1a_2}=-\Gamma^{y}_{yy}\delta_{a_1a_2}=-{1\over{y}}\delta_{a_1a_2}}}
one  easily computes:
\eqn\grav{\eqalign{\nabla_A\nabla^A\Omega_{a_1...a_s}(x)\equiv
(\nabla_{a}\nabla^a+\nabla_y\nabla^y)
\Omega_{a_1...a_s}(x)=(\partial_a\partial^a-\Lambda{s(s+d)})\Omega_{a_1...a_s}}}
Substituting (62) into the  AdS Fronsdal's operator in the momentum space 
(with the Fourier transformed boundary coordinates)
gives
\eqn\grav{\eqalign{({\hat{F}}_{AdS}\Omega(p))^{a_1...a_s}=
-p^2\Omega_m^{a_1...a_{s-1}}(p)
+
{\Sigma_1}(a_1|a_2...a_{s-1})p_tp^{a_1}\Omega_m^{a_2...a_{s-1}t}
\cr
-{1\over2}\Sigma_2(a_{s-2},a_{s-1}|a_1,...a_{s-3})p^{a_{s-1}}p^{a_{s-2}}
(\Omega_m^\prime)^{a_1...a_{s-3}} 
+\Lambda(s+3-d)\Omega_m^{a_1...a_{s-1}}
}}
Thus the $\beta$-functions for the $V_{s-1|0}$ vertex operators 
coincide with $AdS$ Fronsdal operators precisely for $AdS_5$ case ($d=4$).
For other values of $d$ the string theoretic calculation of the mass-like
factor $m_\Omega^2\sim{\Lambda}(s-1)$ is still proportional to $s$, 
but there is a discrepancy
proportional to $d-4$. This discrepancy can always be cured, however, 
by  suitable
modification of the $\psi$-part of the vertex operators of the type (33).
This modification typically involves the shift 
of the canonical picture of the operator
for the Fronsdal's field from $2s-3$ to 
$2s-3+|d-4|$ and is somewhat tedious, but straightforward, 
with the explicit
form depending on $d$. However, the shift 
doesn't change the order of the cohomology,
which is still $H_{s-2}\sim{H_{-s}}$ for each value of $s$.
The Regge-style behavior (57) of the mass-like terms  in 
Fronsdal operators is thus
the consequence of the cohomology structure of the  
higher spin  vertices in the ``larger''
string theory.

\centerline{\bf 5. Conclusions}

We have shown that the massless higher spin operators (33), 
although initially constructed 
around the flat background in $d$ dimensions, 
 lead to the low-energy
higher spin dynamics in the underlying $AdS_{d+1}$ space,
 which presence is initially hinted at
by the hidden symmetries of the RNS action (3), (11), (12)
 and by the cosmological terms appearing
in the $\beta$-function of the spin 2 operator (13) identified with the 
gravitational fluctuations around $AdS$ vacuum. 
In this paper we limited ourselves
to the special case of vertex operators, describing the space-time
higher spin fields polarized (and propagating) along $AdS$ boundary.
The generalization to the bulk case  involves switching on 
the Liouville mode in 
expressions for the operators, which accounts for the radial $AdS$ direction.
This generalization shall be important to perform since 
hopefully it shall reveal 
interesting interplays between $AdS$ geometry  and Liouville central charge 
in various dimensions ~{\selfsw}, 
as well as nontrivial relations between Liouville
structure constants and those of higher spin algebra in various dimensions.
Another important direction to explore is related to the
 higher order corrections 
to the $\beta$-functions of
the higher spin operators, mixing the Weyl transformations with the higher-order
vertex operator contributions in the sigma-model (1). 
One obvious complication
that can be seen immediately is that the 
cohomology argument (53) allowing us to deduce 
the $\beta$-functions for the Fronsdal 
fields in the leading order by studying those for the extra
field operators in the frame-like formalism, is longer  
valid at higher orders, with the contributions
to the $\beta$-functions no longer being linear. 
At the same time, manifest expressions for vertex operators
for Fronsdal's higher fields are generally too complicated 
to work with in a straightforward way,
unless some structural algorithm may be found.
One could still hope though that, with certain modifications 
the cohomology argument (53)
could  still work at higher orders, 
allowing to compute the low-energy couplings
of Fronsdal fields by using the extra field operators 
which structure is far simpler.
We hope to elaborate on the higher order contributions 
in the near future, with the work currently in
progress.
The off-shell arguments considered in this paper strongly suggest
that the most natural string-theoretic
 framework for understanding the structure of the higher spin interactions
at higher  orders is the cubic-like string field theory, 
extended to ghost cohomologies
of higher orders, containing the higher spin operators.
The relevant objects to  compute in such an approach are the 
off-shell correlators
\eqn\lowen{A_N{\sim}<T_{z{\bar{z}}}...T_{z{\bar{z}}}V_{s_1}....V_{s_N}>}
with Vasiliev-type fields being on the worldsheet boundary 
and the Weyl generators
inserted in the bulk. The insertions of 
Weyl generators account for the $AdS$ curvature effects
in higher spin interactions, with the number of the insertions 
corresponding to the order in the cosmological
constant. In general this is not an easy computation to get through, however at 
the first nontrivial order in 
cosmological constant $\Lambda$ (with only one $T$-insertion in SFT 
correlators)
the formalism of Sen-Zwiebach type of open superstring field theory ~{\szw}
(extended to higher cohomologies)
 can hopefully be used,
at least for the fields of Vasiliev's type.
The key point here  is that equations of extended superstring field theory
$\sim{Q\Psi\sim\Psi\star\Psi}$ hold the information about
 higher spin couplings at all orders
similarly to Vasiliev's equations. In fact , the isomorphism between
extended string field theory (SFT) and Vasiliev's equations  
may ultimately be a correct language
to understand  higher spin holography in general. 
Extended string field theory, as we may hope further,
could be an efficient approach
 to understand the dynamics and geometrical aspects 
of multiparticle generalizations and
of quantum higher spin field
theories ~{\vasnew} in general. 
Testing $\beta$-functions for higher spin fields through string field theory
at higher orders, 
to establish their consistency with higher spin 
interactions in $AdS$ should thus provide
a nontrivial check of the conjectured isomorphism
 between equations of Vasiliev and the formalism
of extended SFT.

\centerline{\bf Acknowledgements}

It is a pleasure to  thank Loriano Bonora,
Igor Klebanov, Soo-Jong Rey, Zhenya Skvortsov 
and Misha Vasiliev for useful
comments and discussions.
I also would like to thank the organizers of 
the GGI Workshop on Higher Spin Gauge Theories
at Galileo Institute in Florence for hospitality
where part of this work has been done.
This work was partially supported by the National 
Research Foundation of Korea(NRF) grant funded 
by the Korea government(MEST) through the Center for
 Quantum Spacetime(CQUeST) of 
Sogang University with grant number 2005-0049409.
I alsso acknowledge the support of the NRF grant number 2012-004581.

\listrefs

\end